\documentclass[prl,twocolumn,showpacs,amssymb,superscriptaddress]{revtex4}
\usepackage{graphicx}

\begin{document}

\title{Front Propagation and Diffusion in the A  $\leftrightharpoons$
A $+$ A Hard-core Reaction on a Chain}

\author{Debabrata Panja}\thanks{Present Address: Institute for
Theoretical Physics, Universiteit van Amsterdam, Valckenierstraat 65,
1018 XE Amsterdam, The Netherlands}  \affiliation{Instituut--Lorentz,
Universiteit Leiden, Postbus 9506, 2300 RA Leiden, The Netherlands}
\author{Goutam Tripathy} \affiliation{Institute of Physics,
Sachivalaya Marg, Bhubaneswar - 751005, India}

\author{Wim van Saarloos}  \affiliation{Instituut--Lorentz,
Universiteit Leiden, Postbus 9506, 2300 RA Leiden, The Netherlands}

\date{\today}
\begin{abstract} 
We study front propagation and diffusion in the reaction-diffusion
system A $\leftrightharpoons$ A $+$ A on a lattice. On each lattice
site at most one A particle is allowed at any time.  In this paper, we
analyze the problem in the full range of parameter space, keeping the
discrete nature of the lattice and the particles intact. Our analysis
of the stochastic dynamics  of the foremost occupied lattice site
yields simple expressions  for the front speed and the front diffusion
coefficient  which are in excellent agreement with  simulation results.
\end{abstract}


\maketitle

\section{Introduction}

\begin{figure}
\begin{center}
\includegraphics[width=0.40\textwidth]{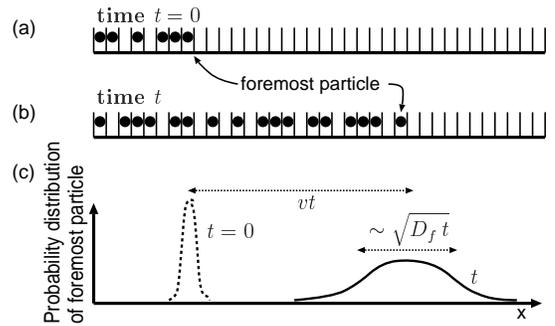}
\end{center}
\caption{{\em (a)} The type of initial condition we consider for our
stochastic model. {\em (b)} Illustration of a typical snapshot of the
state of the system at finite time. The foremost particle has advanced
to the right relative to the one where it started at $t=0$. {\em (c)}
Qualitative sketch of the probability distribution function for the
foremost particle at $t=0$  (dashed line) at large times $t$; the
center of the peak drifts with speed $v$, while the peak widens
proportional to $\sqrt{D_f \,t}$. \label{illus}}
\end{figure}

In this paper, we study the propagation and diffusion of a {\em front
} in the   A $\leftrightharpoons$ A  $+$ A  reaction on a chain, in
the case that there cannot be more than one A-particle  on each
lattice site (``hard-core exclusion''). The front propagation problem
we consider is the following. We start from a situation illustrated in
Fig. \ref{illus}{\em (a)}  in which there are no A-particles at all on
the right half of the system, while there is a nonzero density of
particles on the left.  The object of study is then the asymptotic
average speed $v$ with which the region with a nonzero density of
particles expands to the right, as well as the effective diffusion
coefficient $D_f$ of this ``front''.  For the hard-core exclusion
problem, the front position is most conveniently defined as the
position of the foremost (rightmost) particle, see
Fig. \ref{illus}{\em (a-b)}. The average front speed and front
diffusion coefficient are then the average drift speed $v$ and the
diffusive spreading $\sim \sqrt{D_ft}$ of the width of the probability
distribution $P_{k_f}(t)$ for the location $k_f$ of the foremost
occupied lattice site, as illustrated in Fig. \ref{illus}{\em
(c)}. One of the main results of the paper is a simple  expression for
$v$ and $D_f$, which is accurate in the range where the deviations
from the mean field theory are large. Our results  reduce to an exact
expressions derived before for the particular case in which the
particle diffusion coefficient $D$ and annihilation rate  $W$ are
equal \cite{ba2} and our expression for the front speed $v$ reduces to
the approximate expression obtained for the special case $W=0$ in
\cite{kerstein1,bramson,kerstein2}. In addition, we study the average
particle profile  behind the foremost occupied lattice site and
analyze how its behaviour affects the average front speed and
diffusion.

The perspective of this work lies in the issues that have emerged from
the surprising findings for fronts in this reaction-diffusion system
in the limit in which $N$, the  average number of particles  per
lattice site in equilibrium, is large.  In a lattice model, one can
tune $N$ by allowing more than one particle per site (no hard-core
exclusion) and changing the ratio $k_{\rm b}/k_{\rm d}$, where $k_{\rm
b}$ is the reaction rate for birth processes A $\to $ 2A and $k_{\rm
d} $ the reaction rate for death processes, 2A $\to$ A, as the average
equilibrium number of particles $N= k_{\rm b}/k_{\rm d}$. In the limit
$N$$\to$$\infty$, the normalized particle density $\rho_i \equiv
N_i/N$ then obeys a mean-field equation which is a lattice analogue of
the continuum reaction diffusion equation $\partial_t \rho =
D\partial^2_x \rho + \rho - \rho^2$, where $D$ is the diffusion rate
of individual particles on the chain. The front problem mentioned
above,  i.e., the propagation of a front into an empty region, then
corresponds in the mean-field limit $N $$\to
$$\infty$ to a front propagating into the linearly unstable state
$\rho=0$  (the  mean field behavior is also obtained in the limit  in
which the particle diffusion coefficient $D\to \infty$
\cite{kerstein1,bramson,kerstein2}, but  we will focus on the case in
which the diffusion coefficient is finite and comparable to  the
growth and annihilation rates). The behavior of such fronts in
deterministic continuum equations has been studied since long and is
very well understood (see, e.g., \cite{aronson,ebert}). Since the
nonlinear front solutions are essentially ``pulled along'' by the
growth of the leading edge where $\rho\ll 1$, such fronts are often
referred to as {\em pulled} fronts \cite{ebert}. The remarkable
discovery of the last few years has been that since the propagation is
driven by the region where $\rho$ is small, they are particularly
sensitive to  the discrete nature of the particles which manifests
itself in changes in the dynamics  when $\rho$ becomes of order
$1/N$. Indeed, Brunet and Derrida discovered that the convergence to
the mean-field limit  is extremely slow with $N$:  the average front
speed $v$ converges as $1/\ln^2 N$ to the mean-field value \cite{bd}.
This is in contrast to the fact that for pushed fronts, the
convergence to asymptotic speed behaves as a power of $1/N$.  This
slow convergence has been confirmed for a variety of models
\cite{breuer,bd,bd2,vanzon,kns,levine,PvS1,PvS2}. In addition, in a
model that Brunet and Derrida studied in Ref. \cite{bd2}, the front
diffusion coefficient $D_f$ was numerically shown to vanish only as
$1/\ln^3 N$.

The dominant asymptotic correction to the mean-field result for the
front speed in the limit $N\rightarrow\infty$ traces simply to the
change in the dynamics at $\rho = {\mathcal O} (1/N)$ \cite{bd}, and
as a result appear to be universal. However, all corrections beyond
the asymptotic one appear to depend {\em non-universally\/} on the
detailed stochastic dynamics at the foremost occupied site and those
closely behind it, where asymptotic techniques are of no use since the
number of particles involved in the dynamics is small
\cite{PvS2}. Moreover, the stochastic dynamics in the tip region even
seems to be strongly nonlinearly coupled to the uniformly translating
average front profile behind the tip.

For analyzing these effects for finite values of the particle
diffusion coefficient $D$ and particle number $N$, it is found to be
expedient to develop a stochastic front description by focussing on
the behavior of the foremost particle or the foremost occupied bin
\cite{PvS2}. As it turns out, this idea traces back to the earlier
work by Kerstein \cite{kerstein1,kerstein2} and Bramson and coworkers
\cite{bramson}. These authors analyzed the average front speed $v$ for
a special case of the model we investigate here, namely the case in
which the particle annihilation rate $W=0$. In this case, one can
formulate a self-consistent dynamics for the two foremost particles \cite{eq},
but this important simplification is lost when $W\neq 0$
\cite{neq}.
 Motivated by the desire to understand the ingredients
necessary to analyze the stochastic front behavior for finite values
of $D$, $W$  and $N$, we focus here on analyzing both $v$ and $D_f$ in
the case in which all the transition rates are comparable; our
analysis includes the special point $D=W$ where an exact result was
obtained by ben-Avraham \cite{ba2}.

\section{The Model, Front Speed and Front Diffusion}
 
\begin{figure}
\begin{center}
\includegraphics[width=0.45\textwidth]{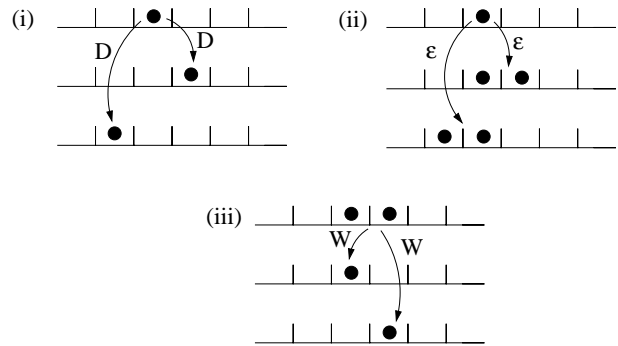}
\end{center}
\caption{The microscopic processes that take place inside the system:
{\em (i)} A  diffusive  hop with rate $D$ to a neighboring empty site;
{\em (ii) } Creation  of a new particle on a site neighboring an
occupied site with rate $\varepsilon$; {\em (iii)} Annihilation  of a
particle on a site adjacent to an occupied site at a rate
$W$. \label{fig1}}
\end{figure}

We now turn to the details of our model and our results for the
stochastic fronts. We consider a chain on which A particles can
undergo the following three basic moves, shown in Fig. \ref{fig1}:
{\em (i) } A particle can diffuse to any one of its neighbour lattice
sites with a diffusion rate $D$, provided this neighboring site is
empty. {\em (ii)} Any particle can give birth to another one on any
one of its empty neighbour lattice site with a birth rate
$\varepsilon$. {\em (iii)} Any one of two A particles belonging to two
neighbouring filled lattice sites can get annihilated with a death
rate $W$.

Note that in the above formulation, diffusive hops to neighboring
sites which are occupied are not allowed. We can also think about
these stochastic moves differently: for example, we can allow nearest
neighbour diffusive hops to a site which is already occupied be
followed by an instantaneous  annihilation of one of the two
particles. If we do so, then the diffusive process contributes to the
annihilation of particles. However, in this paper we shall stick to
the convention that diffusive hops are allowed only to empty sites.

As noted before, earlier work on models of this type includes that of
Kerstein \cite{kerstein1,kerstein2} and Bramson {\em et al.}
\cite{bramson} on the case $W=0$ and that of ben-Avraham on the case
$D=W$ \cite{ba2} (also, variants of this model have been analyzed in
Refs. \cite{ba1,doe1,doe2}). Notice that  in the general case there
are essentially only two nontrivial parameters in the model, e.g. the
ratios $D/\varepsilon$ and $D/W$, since an overall multiplicative
factor just sets the time scale.  Our interest is in the parameter
range where both of these ratios are ${\cal O}(1)$; when these ratios
tend to infinity, the front speed approaches the mean field value
\cite{kerstein1,bramson,kerstein2}.

For an ensemble of front realizations, let us denote the probability
distribution for the foremost occupied lattice site to be at lattice
site ${k_f}$ by $P_{k_f}(t)$. The evolution of $P_{k_f}(t)$ is then
described by
\begin{eqnarray}
\frac{dP_{k_f}}{dt}\,=\,(D\,+\,\varepsilon)\,P_{{k_f}-1}\,+\,\left[D\,P^{\mbox{\scriptsize
empty}}_{{k_f}+1}\,+\,W\,P^{\mbox{\scriptsize
occ}}_{{k_f}+1}\right]\nonumber\\&&\hspace{-6.9cm}-\,(D\,+\,\varepsilon)\,P_{{k_f}}\,-\,\left[D\,P^{\mbox{\scriptsize
empty}}_{{k_f}}\,+\,W\,P^{\mbox{\scriptsize occ}}_{{k_f}}\right]\,.
\label{e2}
\end{eqnarray}
Here $P^{\mbox{\scriptsize occ}}_{{k_f}}(t)$ and $P^{\mbox{\scriptsize
empty}}_{{k_f}}(t)$  denote the joint probabilities that the foremost
particle is at site $k_f$ and that the site  $k_f-1$ is occupied or
empty, respectively.  Clearly, $P_{k_f}(t)=P^{\mbox{\scriptsize
occ}}_{{k_f}}(t)+P^{\mbox{\scriptsize empty}}_{{k_f}}(t)$, and
$\sum_{k_f}P_{k_f}(t)=1$. The first term on the r.h.s. of
Eq. (\ref{e2}) describes the increase in $P_{k_f}(t)$ due to the
advancement of a foremost occupied lattice site from position
${k_f}$$-$$1$, while the second term describes the increase in
$P_{k_f}(t)$ due to the retreat of a foremost occupied lattice site
from position ${k_f}$$+$$1$. The  third and the fourth terms,
respectively, describe the decrease in $P_{k_f}(t)$ due to the
advancement and retreat of a foremost occupied lattice site from
position ${k_f}$.

From the definition of $P_{k_f}(t)$, the mean position and the width
of the distribution for the positions of the foremost occupied lattice
sites are defined  as $x(t)=\sum_{k_f}{k_f}P_{k_f}(t)$ and
$\langle\Delta x^2(t)\rangle=\sum_{k_f}[{k_f}-x(t)]^2P_{k_f}(t)$
\cite{notedeltat}. The mean speed and  diffusion coefficient of the
front are thus given in terms of these quantities as  the
$t$$\to$$\infty $ limit of $v=d x(t)/dt $ and $\langle\Delta
x^2(t)\rangle = 2D_f t$ --- see Fig. \ref{illus}{\em (c)}.  To obtain
them, we need the expressions of $P^{\mbox{\scriptsize
occ}}_{{k_f}}(t)$ and $P^{\mbox{\scriptsize empty}}_{{k_f}}(t)$. To
start with, we have
\begin{equation}
P^{\mbox{\scriptsize occ}}_{k_f}(t)=\rho_{k_f-1} P_{k_f}(t),
\label{neweq}
\end{equation}
where $\rho_{k_f-1}$ is the conditional probability of having the
$({k_f}$$-$$1)$th lattice site occupied (the foremost particle is  at
the ${k_f}$th lattice site). The set of conditional occupation
densities $\rho_{k_f -m}$ for  $m \ge 1$ can be thought of as
determining the front profile in a frame moving with each front
realization. For obtaining $v$ and $D_f$, we simply need to know the
asymptotic long-time limit  $\rho_{k_f-1}( t\to\infty)$, which from
here on we will denote simply as $\rho_{k_f-1}$. Given $\rho_{k_f-1}$,
it is then straightforward to obtain from Eq. (\ref{e2}) and the
conditions $P_{k_f}(t)=P^{\mbox{\scriptsize
occ}}_{{k_f}}(t)+P^{\mbox{\scriptsize empty}}_{{k_f}}(t)$ and
$\sum_{k_f}P_{k_f}(t)=1$
\begin{eqnarray}
v\,=\,\frac{dx}{dt}\,=\,\varepsilon\,-\,\rho_{k_f-1}(W\,-\,D)\quad\quad\mbox{and}\nonumber\\&&\hspace{-6.15cm}\frac{d\langle\Delta
x^2\rangle}{dt}\,=\,2D\,+\,\varepsilon\,+\,\rho_{k_f-1}(W\,-\,D)\,.
\label{e3}
\end{eqnarray}
Of these, the second equation indicates that the front wandering is
diffusive, and an expression of the front diffusion coefficient $D_f$
is therefore given by
\begin{eqnarray}
D_f\,=\,\frac{1}{2}\left[2D\,+\,\varepsilon\,+\,\rho_{k_f-1}
(W\,-\,D)\right]\, .
\label{e4}
\end{eqnarray}
As noted already by ben-Avraham \cite{ba2} in a continuum formulation
of the present model, for the special case  $D=W$ the unknown quantity
$\rho_{k_f-1}$ drops out of Eq. (\ref{e3}); it thus leads to the {\em
exact} results $v=\varepsilon$ and $D_f=D+\varepsilon/2$ as a  special
cases of Eq. (\ref{e4}) for {\bf $D=W$}.  We also note that if we use
Eq. (\ref{neweq}) in Eq. (\ref{e2}),  the latter equation has the form
of the master equation for a single  random walker on a chain. Thus we
can think of the foremost particle as executing a biased random walk,
and $D_f$ as the effective diffusion coefficient of this
walker. Moreover, if we eliminate $\rho_{k_f-1}$ from Eqs. (\ref{e3})
and (\ref{e4}), we  get the following {\it exact\/} relation
\begin{equation}
v/2\,+\,D_f\,=\,D\,+\,\varepsilon\,. \label{exact}
\end{equation} 

In order to obtain an explicit prediction for $v$ and $D_f$, we need
an expression for $\rho_{k_f-1}$. Far behind the front the particle
density will approach the homogeneous equilibrium density
$\overline{\rho}$: $\lim_{m\to\infty} \rho_{k_f-m} = \overline{\rho}$.
From the  master equation it is easy to show that the  homogeneous
equilibrium solution for the total probability is of product form (so
that the probability of having different sites is occupied is
uncorrelated), and that the equilibrium occupation  density
$\overline{\rho}$ is simply given by
$\overline{\rho}=\varepsilon/(\varepsilon+W)$.

The crudest approximation for the front profile $\rho_{k_f-m}$ and in
particular for $\rho_{k_f-1}$ is to just  take $\rho_{k_f-1} \approx
\overline{\rho}$. Substitution of this approximation into
Eqs. (\ref{e3}) and (\ref{e4}) immediately yields our main result,
\begin{eqnarray}
v\,=\,\frac{\varepsilon\,(\varepsilon\,+\,D)}{\varepsilon\,+\,W}\quad\mbox{and}\quad
D_f\,=\,\frac{(\varepsilon\,+\,2W)(D\,+\,\varepsilon)}{2(\varepsilon\,+\,W)}\,.
\label{e5}
\end{eqnarray}
For $W=0$, the expression for $v$ reduces to the one obtained in
\cite{kerstein1,bramson,kerstein2}.

In what follows, we will first compare these approximate expressions
for $v$ and $D_f$ to the results of computer simulation for the case
$D/\varepsilon=1$ , and then investigate the appropriateness and
shortcomings of the approximation $\rho_{k_f-1}\approx
\overline{\rho}$.

\begin{figure}
\begin{center}
\includegraphics[width=0.37\textwidth,angle=270]{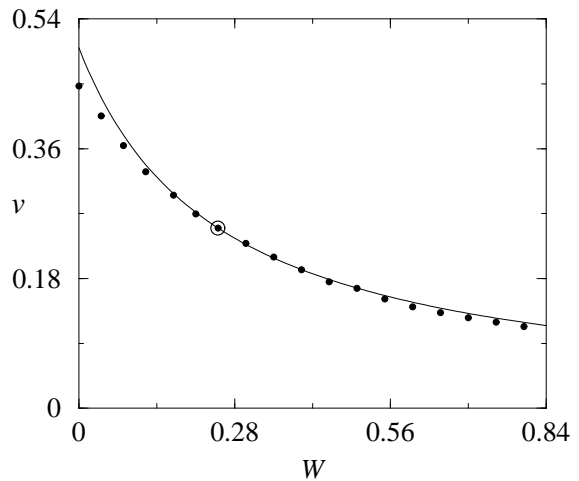}
\end{center}
\caption{Comparison of the expression of $v$ in Eq. (\ref{e5}) (solid
line) with stochastic simulation data (filled circles), for
$D=\varepsilon=0.25$. The error in the data is of the order of the
size of the symbols. The corresponding data point for $D=W$, as
analyzed in Ref. {\protect\cite{ba2}} is shown by the larger open
circle. \label{fig2}}
\end{figure}

\begin{figure}
\begin{center} 
\includegraphics[width=0.37\textwidth,angle=270]{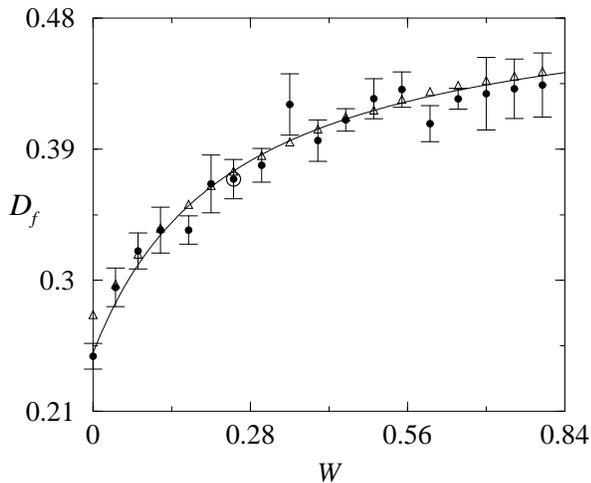}
\end{center} 
\caption{Comparison of the front diffusion coefficient according to
Eq. (\ref{e5}) (solid line) with stochastic spreading data (filled
circles) and with Eq. (\ref{exact}) (open triangles), for
$D=\varepsilon=0.25$. The large open circle once again corresponds to
the direct measurement of the effective front diffusion coefficient
for $D=W$, as analyzed in Ref. {\protect\cite{ba2}}.
\label{fig3}} 
\end{figure}

The comparison of Eq. (\ref{e5}) with stochastic simulation data for
$D=\varepsilon=0.25$ are presented in Figs. \ref{fig2} and \ref{fig3}
as a function of $W$ for $D=\varepsilon=0.25$.  The simulation
algorithm has been adopted from \cite{PvS2}, and is essentially the
same one as in \cite{breuer}. The speed $v$ has been obtained directly
from the average position of the foremost occupied lattice site in a
single long run according to $v(t) = [x(t)-x(t_0)]/(t-t_0)$
corresponding to $x(t)-x(t_0)=15000$ consecutive forward jumps. The
diffusion coefficient has been determined both from the speed
measurements via (\ref{exact}) and from data for the average diffusive
spreading during 1000 time intervals $\Delta t$ up to 500 taken from 5
long runs (of which the data from the first 5000 consecutive forward
jumps of the foremost occupied lattice site were ignored, so as to
eliminate initial transient effects).  For each of these runs, the
mean square displacement $\langle\Delta x^2\rangle$ was confirmed to
grow linearly with time.  Figures \ref{fig2} and \ref{fig3} show that
our approximate expressions (\ref{e5}) for the speed and diffusion
coefficient (solid line) are quite accurate for $D/\varepsilon=1$ over
the whole range of values of $W$ where we have performed simulations;
interestingly, the values of $D_f$ obtained from the speed
measurements via Eq. (\ref{exact}) are more accurate than those
obtained directly from the diffusive spreading. The error bars in Fig.
\ref{fig3} correspond to the standard deviations of $D_f$ values
obtained from 5 long runs.

\begin{figure}
\begin{center}
\includegraphics[width=0.37\textwidth,angle=270]{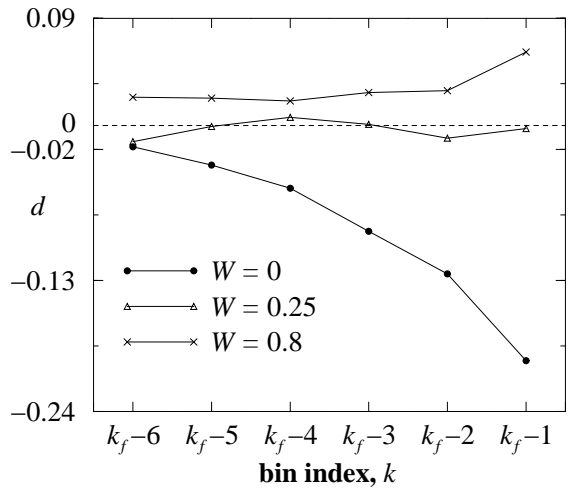}
\end{center}
\caption{Relative  deviation
$d=(\rho_k-\overline{\rho})/\overline{\rho}$ of the average density
from $\overline{\rho}= \varepsilon/(\varepsilon+W)$ for the first six
lattice sites to the left of the foremost occupied lattice site,
$k_f$, for $D=\varepsilon=0.25$ and three different values of $W$.
\label{fig4}}
\end{figure}

We now return to the issue of the appropriateness of the assumption
$\rho_{k_f-1}=\overline{\rho}$. While the agreement between the
theoretical prediction for $v$ and $D_f$ gives empirical evidence that
this assumption is a reasonably good one, we see from Fig. \ref{fig2}
that although Eq. (\ref{e5}) agrees well with the simulation data,
there are small but systematic deviations on both sides of this
region. These deviations can be explained as follows:  As $W\to 0$,
$\overline{\rho}\uparrow 1$: far behind the front all lattice sites
are occupied. However, the density of particles just behind the
foremost one is smaller, since it takes a finite time for the density
to relax to the asymptotic one. For large values of $W$,  the
effective diffusion rate is much larger than the drift rate, as
Eq. (\ref{e5}) shows. As a result, once again the density of particles
just behind the foremost one also has relatively small time to relax
to the asymptotic value. This is reflected in the difference between
$\rho_{k_f-1}$ and $\overline{\rho}$ in Fig. \ref{fig4}.

The above trends are borne out by the simulation results of
Fig. \ref{fig4}, where we plot the relative  deviation
$d=(\rho_k-\overline{\rho})/\overline{\rho}$ for
$k=k_f-1,\ldots,k_f-6$. First of all, the data confirm  that unless
the value $W$ is too small, $\rho_{k_f-1}=\overline{\rho}$ is quite a
good approximation, and that the density behind the foremost particle
is enhanced for large $W$ and reduced for small $W$. We also note that
we have verified that  if one substitutes the $\rho_{k_f-1}$ values
for $W=0$ and $W=0.8$ from Fig. \ref{fig4} into Eq. (\ref{e2}), one
does recover the corresponding measured speeds, as one should.

\section{conclusion}

In conclusion, this work clearly illustrates that the concept of the
dynamics of the foremost occupied lattice site, in Refs.
\cite{PvS2,kerstein1,bramson,kerstein2} and here, can be a  viable
route towards analyzing the front propagation and diffusion in
stochastic lattice models. In the present $N\le 1$ model a simple
approximation for  the interaction of the foremost particle with the
front region behind it already yields quite accurate results for $v$
and $D_f$. We hope that this success provides new motivation and
inspiration to tackle the complicated case in which $N$ is large but
finite. 

In principle, it should be possible to extend the analysis in the
spirit of the one developed by Kerstein \cite{kerstein1,kerstein2} to
get successively more accurate expressions for $\rho_{k_f-1}$, and
correspondingly for the front speed and diffusion coefficient. In
particular, such extensions might allow one to use the results in a
wider parameter range, such as $D/W\rightarrow\infty$ while
$D/\varepsilon\sim{\cal O}(1)$, or $D/\varepsilon\rightarrow\infty$
while $W/\varepsilon\sim{\cal O}(1)$. However, inspection of  the
earlier analysis suggests that such higher order analytical
expressions of $\rho_{k_f-1}$ are less trivial to obtain than one
might expect at first sight. More precisely, in the light of
\cite{eq,neq}, it is  clear that for $W\neq 0$, the master equation
for the probability that the two foremost particles are separated by
$k$ lattice sites couples to probability distributions
involving particles that are further back. While it is certainly
possible to solve the master equation numerically, it does not appear
to lead one to an analytical expression of $\rho_{k_f-1}$ that
provides a better approximation than what we have used in this paper.

The work by D. P. and that of G.T. during an earlier stay at
Universiteit Leiden was  supported by the Foundation FOM
(``Fundamenteel Onderzoek der Materie'').


\begin{thebibliography}{02}
\bibitem{eq} As can be seen from Eqs. (\ref{e3}), to obtain the front
speed and the front diffusion coefficient, one needs the expression of
$\rho_{k_f-1}$, the probability of the occupancy of the lattice site
just behind the foremost particle. For $W=0$, the ``two-particle
self-consistent'' approach developed in
Refs. \cite{kerstein1,kerstein2} can obtain a better approximation of
$\rho_{k_f-1}$ than what we present in this paper, from the solution
of the master equation for the probability that the two foremost
particles are separated by $k$ lattice sites.
\bibitem{neq} For $W=0$, the master equation for the probability that
the two foremost particles are separated by $k$ lattice sites can be
closed in a simple manner \cite{kerstein1,kerstein2}. In this
formalism, no particle gets annihilated, and as a result, the
hierarchy of equations for the joint probability density distribution
of the two foremost particles can be closed easily at the simplest
level, since in the absence of annihilation, the third foremost
particle never becomes the second foremost particle. At this level,
the expression of $\rho_{k_f-1}$ can then be analytically solved,
leading to a better approximation than what we use in this paper for
$W=0$. Of course, the master equation can be closed at a higher level,
by considering more than two foremost particles to determine
$\rho_{k_f-1}$, but then one does not obtain an analytical expression
of $\rho_{k_f-1}$. As soon as $W\neq 0$, this is not true anymore:
consider the following situation where the two foremost particles are
next to each other. With annihilation of particles allowed, one of
them can annihilate the other, and then the probability distribution
function of the two foremost particles is crucially coupled to those
which involve particles further back at the simplest level.
\bibitem{notedeltat} One can also define the foremost occupied lattice
site for a realization as the one on the right of which no lattice
site has ever been occupied before, and obtain the front speed from
this definition following \cite{PvS2}. Both of them of course yield
the same result due to  the $t\rightarrow\infty$ limit.
\bibitem{noteother} We note here that the choice of $D/\varepsilon=1$
is only coincidental, and the relative agreements in Figs. \ref{fig2}
and \ref{fig3}, are not affected so long as $D$ and $\varepsilon$ are
of the same order of magnitudes.
\end{thebibliography}
\end{document}